# Quantitative evaluation of cracks' depth on a thin aluminum plate by using eddy current pulse-compression thermography

*Abstract*— Eddy current stimulated thermography is an emerging technique for non-destructive testing and evaluation of conductive materials. However, quantitative estimation of the depth of subsurface defects in metallic materials by thermography techniques remains challenging due to significant lateral thermal diffusion. This work presents the application of eddy current pulse-compression thermography to detect surface and subsurface defects with various depths in an aluminum sample. Kernel Principal Component analysis and Low Rank Sparse modelling were used to enhance the defective area, and cross-point feature was exploited to quantitatively evaluate the defects' depth. Based on experimental results, it is shown that the crossing point feature has a monotonic relationship with surface and subsurface defects' depth, and it can also indicate whether the defect is within or beyond the eddy current skin depth. In addition, the comparison study between aluminum and composites in terms of impulse response and proposed features are also presented.

*Index Terms*— Defect depth evaluation, Crossing-Point Feature, Eddy Current Pulse-Compression Thermography, Kernel Principal Component Analysis, Low-Rank Sparse modelling

## I. INTRODUCTION

Aluminum (Al) material is used in many industrial applications including aerospace components due to its low weight and high strength-to-weight ratio. Therefore, the evaluation of surface and subsurface cracks and discontinuities in Al components is of high importance. So far, various Active Thermography (AT) techniques have been exploited to evaluate the defect depth in conductive metallic materials including Pulsed Thermography (PT) [1] , Pulse-Phase Thermography (PPT) [2] and Lock-in Thermography (LT) [3]. These techniques can be applied on most of the heat sources, either being surface stimulation, *e.g.* flash lamp [4] and LED [5], or volumetric *e.g.* eddy current.

Among these AT techniques, one of the most extensively applied is the Eddy Current Pulsed Thermography (ECPT). ECPT uses a coil driven by an alternating current to generate Eddy Current (EC) inside the Sample Under Test (SUT), thus increasing the temperature of the SUT due to the Joule effect. ECPT has been demonstrated being able to detect surface cracks with higher reliability and reproducibility than vibro-thermography and laser thermography [6]. In addition, the high performance of ECPT, *e.g.* robustness to lift-off variations and applicability to defect orientation characterization and depth estimation, makes it suitable for fast quantitative evaluation [7]. The application of ECPT has been investigated for both detection and characterization of material degradation and failure such as fatigue cracks, corrosion and residual stress [8].

The detection of defects in metallic materials by means of ECPT relies on two physical phenomena, since surface and subsurface defects can indeed react differently to the excitation: Joule heating through eddy current and the heat diffusion. The former plays the main role in the detection of surface and subsurface cracks located within the range of the theoretical skin depth of EC, due to increased current density. Therefore, features in the heating stage (*e.g.* peak time, peak magnitude) are investigated to characterize surface or shallow sub-surface defects. Instead, when the defect is located beyond the skin depth, the detection is mainly due to the diffusion of Joule heating from outer areas to inner ones. In this case, relevant features that are useful for classification purposes are extracted mainly during the cooling stage (*e.g.* phase information, thermal signal reconstruction). However, due to the low thickness and high thermal conductivity of SUT, it is more challenging to detect subsurface defects beyond the skin depth since the lateral heat diffusion is not negligible with respect to the through-thickness one.

To tackle this challenge and to improve the detection capability of standard ECPT, a combination of pulse-compression (PuC) techniques and EC excitation was proposed recently as Eddy Current Pulse-Compression Thermography (ECPuCT) scheme, either by means of coded excitations [9] or without using them [10]. It has been reported in literature that PuC combined with AT, improves the achievable Signal-to-Noise Ratio (SNR) even while using low-power heat sources [11, 12]. The present authors recently showed how ECPuCT can be fruitfully applied on the inspection of CFRP material [9]. This technique is based on the exploitation of a phase-modulated current waveform to excite EC within the sample and of a proper filter matched to this excitation (the so called matched-filter) that is applied pixel-wise on time trends to retrieve the impulse response of the SUT as a series of thermograms. In addition, in [9] such approach was further extended to quantitative analysis by exploiting proposed feature extraction strategy.

Besides improving the detectability of defects beyond the skin depth by ECPuCT technique, the application of feature extraction methods on thermogram time sequences, i.e. on impulse responses, is equally important to conduct quantitative study of defect depth. Feature extraction methods can be divided into two main categories: those analyzing the transient response of each single pixel of thermogram, such as peak time, response differential, Fourier transform , wavelet transform [13], and the ones processing thermal images to extract image-based features for estimation of defect location, such as Principal Component Analysis (PCA), Independent Component Analysis ICA and Singular Value Decomposition (SVD) [14]. These two approaches have pros and cons: transient response-based features can directly quantify signals but lack of visualization ability to provide the profile of defect, while image-based features on thermal images can provide the defect profile based on statistical distribution. However, the results of the latter approach are difficult to be interpreted physically, hindering the further application of quantitative analysis.

To solve the mentioned problem, this paper provides two state-of-the-art pattern recognition techniques, Kernel Principal Component Analysis (K-PCA) and Low Rank Sparse pattern modelling (LRS), to directly decompose impulse responses obtained by ECPuCT rather than thermal images, according to second approach, for subsurface and surface defect pattern defection.

Overall, this work presents novel approaches on experiment, post-processing and quantitative evaluation of subsurface and surface defects in an artificially-defected thin Al plate, of which the flow diagram is shown in Fig. 1. As showed in Fig. 1-block 1, a Barker Code (BC) modulated EC excitation was applied to obtain raw data of subsurface and surface defects with various depths. Then, PuC process was implemented on the raw data to retrieve the impulse response of each individual pixel as shown in block 2. In block 3, two defect detection techniques, K-PCA and LRS, were used to decompose the impulse responses for defective area location. A comparison between these two techniques was also done in terms of SNR values for surface and subsurface detects. Finally, in block 4, features including the crossing point of impulse responses collected on defective and non-defective areas for subsurface and surface defects were used for comparing and evaluating the various defects and their depths. The comparison between cross-point feature for quantitative evaluation of Al and composite samples is also provided in block 4, based on previous work [9].

This paper is organized in five sections: Section II describes the proposed methodologies, including theory of ECPuCT and mathematical implementation of K-PCA and LRS on impulse responses, Section III illustrates the experimental setup and the SUT. In Section IV, the experimental results are presented and discussed, and finally, Section V draws the conclusion and describe the future work.

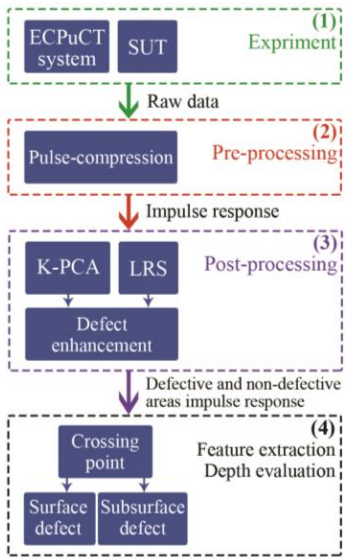

Fig. 1. System diagram of the propose methodology.

## II. Proposed Methodology.

### A. Eddy current pulse compression thermography

Pulse-compression is a wide-spread measurement technique used to estimate the impulse response of a Linear Time Invariant (LTI) system in a noisy environment, or in case of poor SNR values [15]. In standard ECPT, a short pulse excitation induces the eddy currents inside the SUT to heat it up. Please note that "short" is here referred to the typical times of heat propagation. Therefore, the so-provided heating stimulus can be modelled, from a thermal point of view, as a Dirac's Delta function $\delta(t)$, and the corresponding output $y(t)$, i.e. the pixel temperature/emissivity amplitude recorded while time elapses, is a good approximation of the impulse response $h(t)$. Features of interest are obtained by analyzing the $h(t)$ within a chosen range of interest $T_h$, as depicted in Fig. 2. As showed in Fig. 2, in ECPT the excitation is considered instantaneous and the sample's thermal impulse response is measured for a time of interest $T_h$, which is the expected impulse response time duration. In ECPuCT, the sample is excited with a coded excitation of duration $T$ and thermograms are collected for an overall time duration of $T + T_h$. An estimated impulse response $\tilde{h}(t)$ of duration $T_h$ is retrieved after performing the PuC algorithm.

PuC relies on the existence of a pair of signals, an excitation signal $s(t)$ of duration $T$ and bandwidth $B$, and a matched filter $\Psi(t)$, such that their convolution "$*$" approximates the Dirac's Delta function $\delta(t)$:

$$s(t) * \Psi(t) = \tilde{\delta}(t) \approx \delta(t) \qquad (1)$$

If Eq. (1) is satisfied, and $s(t)$ excites a SUT characterized by the ideal impulse response $h(t)$, then an estimate $\tilde{h}(t)$ of the $h(t)$ is obtained by convolving the output signal $y(t)$ with $\Psi(t)$. Eq. (2) shows the mathematical formulation for a single pixel of the acquired thermograms, in presence of an Additive-White-Gaussian-Noise $e(t)$, which is considered uncorrelated with $\Psi(t)$. The impulse response can be obtained as:

$$\begin{aligned}\tilde{h}(t) = y(t) * \Psi(t) &\xrightarrow{y(t)=h(t)*s(t)+e(t)} \\ &= h(t) * s(t) * \Psi(t) + e(t) * \Psi(t) \\ &= h(t) * \tilde{\delta}(t) + \tilde{e}(t) \approx h(t) + \tilde{e}(t)\end{aligned} \qquad (2)$$

The main advantage of ECPuCT with respect to ECPT is that the impulse response can be estimated by delivering energy to the system in a significantly longer time. This flexibility can be exploited, as in the present case, either to increase the SNR by providing more energy than in the pulsed scheme or to use lower power heating sources while maintaining the same SNR level. This latter application is particularly useful to avoid possible thermal shocks in sensitive materials [16]. In both cases, having fixed the heat source power and the excited bandwidth $B$, the SNR gain is proportional to the $T \times B$ product of the coded signal and it can be enhanced almost arbitrarily by increasing the time duration $T$. It should be also noted that the limited $T \times B$ product of practically-employed coded signals results in an $\tilde{h}(t)$ always affected by the so-called "side-lobes", i.e. any local maxima of the $\tilde{\delta}(t)$ in (2) different from the main correlation peak. This can be improved by a proper choice of the matched filter signal $\Psi(t)$ [16]. In this paper,

$s(t)$ is a Barker Code (BC) of order equal to 13 and the matched filter $\Psi(t)$ is the time-reversed sequence of the input coded signal, *i.e.* $s(-t)$ [17]. Details about the BC signal and how to implement it in ECPuCT will be shown in the next section.

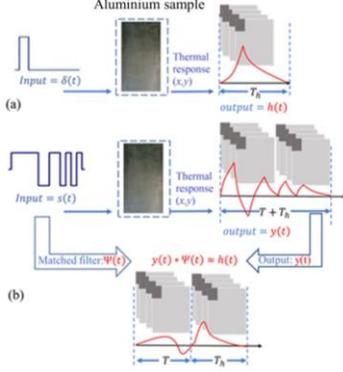

Fig. 2. Comparison of principles between (a) eddy current pulse thermography, (b) eddy current pulse compression thermography.

### B. Defected area detection techniques based on the retrieved impulse response

To quantitatively evaluate the depth of subsurface defects, defected area should be enhanced in order to select the defective pixels for further analysis. The observation model of the recorded thermogram sequences after PuC can be considered as:

$$Y(t) \approx \sum_{i=1}^{C} m_i H_i(t) \quad (3)$$

where $\{H_i(t) \epsilon R^{N_x \times N_y}\}$ represents the set of $C$ patterns, and $N_x \times N_y$ denotes the total number of $x$ and $y$ pixels of the acquired thermograms. These patterns include defective, non-defective, coil and non-heated areas. $m_i$ denotes the mixing parameter that describes the contribution of different thermal patterns to the observation output $Y(t)$. The goal is to extract $H_d(t)$, which describes the defective area impulse response. To accomplish this, two approaches have been exploited here: K-PCA and LRS. Detailed implementation process is shown in Fig. 3 and mathematical explanations are as follows:

#### 1) Kernel principal component analysis

$Y(t)$ introduced in Eq. (3), can be expressed as a combination of every pixel's impulse response as follows:

$$Y(t) = [\tilde{h}_1(t), \tilde{h}_2(t), \ldots, \tilde{h}_{M-1}(t), \tilde{h}_M(t)] \quad (4)$$

The reshaped data can be recognized as a matrix having dimension of $Q \times M$, where $Q$ denotes the number of frames recorded by the IR camera, *i.e.* $T_h \times FPS$ and $M$ is equal to $N_x \times N_y$. To simplify the mathematical explanation, $\tilde{h}(t)$ is here replaced by $\tilde{h}$. By using the kernel method, the impulse response is projected to the kernel space $\phi$, thus obtaining the kernel matrix $K(i,j)$ as:

$$K(i,j) = \frac{1}{M}\sum_{i=1}^{M}\left(\phi(\tilde{h}_i) - \frac{1}{M}\sum_{j=1}^{M}\phi(\tilde{h}_j)\right)\left(\phi(\tilde{h}_i) - \frac{1}{M}\sum_{j=1}^{M}\phi(\tilde{h}_j)\right)^T \quad (5)$$

where $\phi$ is Gaussian kernel function, defined as Eq. (6):

$$\phi(\tilde{h}_i) = \exp\left(-\frac{\|\tilde{h}_i \times \tilde{h}_i^T\|_2^2}{2\sigma^2}\right) \quad (6)$$

The kernel matrix $K(i,j)$ of Eq. (5) can be simply named as $K$. The eigenvector $\alpha$ of $K$ can be obtained as:

$$\lambda_i \alpha_i = K\alpha_i \quad (7)$$

Based on the obtained eigenvectors $\alpha_i$, the enhanced thermal pattern can be projected as:

$$H_d(t) = [\alpha_1, \ldots, \alpha_T] Y(t)^T \quad (8)$$

#### 2) Low rank sparse pattern modelling

Sparse pattern refers to the significance of the data, which can be equalized to defective impulse response. Thus, the observation model $Y(t)$ can be expressed as:

$$Y(t) = [\underbrace{\sum_{i=1}^{C-1} m_i H_i(t)}_{L}] + \underbrace{m_j H_j(t)}_{S} + N \quad (9)$$

where $Y(t)$ is considered as linear combination of three types of $H(t)$'s, which denote the low-rank matrix $L$ (*e.g.* non-defective area, background and non-heated area), the sparse pattern $S$ (*e.g.* defective area impulse response), which contain few non-zero values and the noise $N$. To make it simplified, low rank pattern is denoted by $L$ and sparse pattern is denoted by $S$. To obtain sparse pattern $S$, it can be reformulated as follows:

$$S = MN^T \quad (10)$$

Thus, the extraction of $S$ can be solved as:

$$\min_{M,N,L}\{p\,rank(L) + \varphi_m\|M\|_2^2 + \varphi_n\|M\|_2^2 + \|Y - L - S\|_F^2\} \quad (11)$$

where $p$ controls the rank of $L$, $\varphi_m$ and $\varphi_n$ are for the regulation of $M$ and $N$. The solution of Eq. (11) can be divided into two sub-problems, which are implemented as follows:

$$(L)^i = \arg\min_L\{\|L - (Y - S)^{i-1}\|_F^2 + p\|L\|_*\} \quad (12)$$

$$(S)^i = \arg\min_{M,N}\{\|(Y - (L)^i - S\|_F^2 + \frac{\varphi_m}{2}\|M\|_2^2 + \frac{\varphi_n}{2}\|M\|_2^2\} \quad (13)$$

Eq. (12) is referred as low rank estimation and can be solved by singular value threshold algorithm and Eq.(13) is referred as sparse decomposition and can be solved using Bayesian matrix factorization approach [18]. Once obtained $S$, which contains the defective impulse responses, the damaged area can be enhanced as follows:

$$H_d(t) = S * Y(t)^T \quad (14)$$

These two techniques are applied on the impulse responses and help to identify the defective area for further analysis. To compare the performance of two post-processing techniques, the SNR value is calculated for quantitative study as defined below:

$$SNR(t) = \frac{h_D(t) - \overline{h}(t)}{\sigma_h(t)} \quad (15)$$

where $h_D(t)$ is the impulse response of defected area averaged over a 2×3-pixel region, $\overline{h}(t)$ is the impulse response averaged over all thermogram pixels and $\sigma_h(t)$ is the standard deviation of the same 2×3-pixel region.

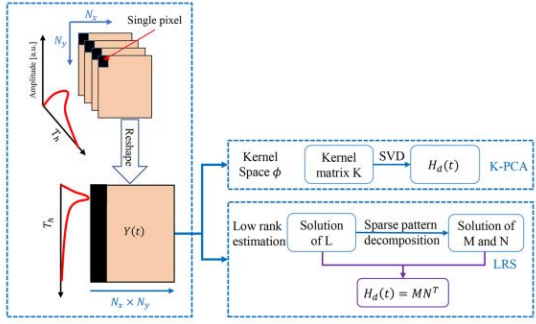

Fig. 3. Implementation of K-PCA and LRS for defect area detection.

## III. Experimental Setup

This section describes the experimental setup and how to employ the coded signal for modulating the induction heating system in on/off state by means of a bipolar BC signal with total bit length of 13. The BC code is not employed in its original bit version, meaning that each "1" and "-1" of the original BC is padded with a series of "1" or "-1" respectively to allow the heat source spreading enough energy toward the SUT. In addition, by changing the single bit duration, one can tune the frequency spectrum of the resulting BC. This assures exciting a continuous range of thermal diffusion length $\mu$ values, which allows defects buried at different depths to be detected. Fig. 4 shows the employed BC signal with bit length of one second at 50 frames per second (FPS). For the chosen BC, the heat emission has an almost flat spectrum from DC to 0.50 Hz, *i.e.* thermal waves are generated in this frequency range having the same amplitude/power density. In ECPuCT, the chosen BC modulates the induction heating unit, *i.e.* the on/off time instants at which a current $I$ of given amplitude $Amp$ and frequency $f_{carrier}$ flows within the coil.

Fig. 5 shows the ECPuCT system. An Agilent 33500B signal generator was used to send both the BC modulating signal to the induction heating system and a reference clock trigger to the IR camera to acquire thermograms at 50 FPS. A Cheltenham EasyHeat 224 induction heating unit is used for exciting the coil with a maximum excitation power and current values of 2.40 kW and 400 A respectively with tunable $f_{carrier}$ within 150 to 400 kHz. For the reported experimental results, values of excitation current $I$ equal to 250 A and $f_{carrier}$ equal to 270 kHz were selected. Water cooling was implemented to cool down the coil and the lift-off maintained at 1.00 mm from the SUT. Only one side of the coil with the length 9.30 mm was selected as linear coil to induce parallel eddy currents inside the SUT. IR camera was the FLIR SC655, equipped with an un-cooled microbolometer detector array with the resolution of 640×480 pixels, the spectral range of 7.5 - 14.0 µm and NETD < 30 mK The IR camera recorded the surface temperature distribution of the $T = 13\ s$ BC as well as additional $T_h = 30\ s$ of cooling period [19] with a frame rate of 50 Hz, see Fig.4 and Fig.6(a). Finally, the captured thermogram sequences were transmitted to a PC for visualization and postprocessing, including signal pre-process, PuC, defect detection and feature extraction. An example of a captured thermal raw signal from the investigated SUT is shown in Fig.6(b) for a single pixel, whilst Fig.6(d) depicts the same signal but after applying PuC, thus after exploiting Eq.(2). For the sake of completeness, it must be note that a step-heating contribution \must be removed from the raw signal before applying faithfully the PuC algorithm, see Fig.6(c). Note that this is an unavoidable passage whenever PuC is aimed at being exploited in combination with unipolar heating sources, as for the present case. The reader is referred for example to [9-12,15,16,19] for understanding how to successfully remove the step-heating contribution - thus to pass from the signal depicted in Fig.6(b) to the one in Fig.6(c), exploiting for example a linear/non-linear fitting function - before applying the PuC algorithm.

Measurements were executed on a specimen with artificial defects. The specimen was made of 2024-T3 aluminum alloy with an electric conductivity equal to 18.8 MS/m, magnetic permeability of 1.26 H/m, and a thickness of 2.00 mm. All the defects were machined as small notches having a length of 3.00 mm and width of 0.10 mm, with varying depths from the inspection surface. Defects' depth varied from 1.60 mm to 0.20 mm with a step of 0.20 mm corresponding to defects D1 to D8 respectively. D9 is a through-hole notch. To have optimal interaction of the induced field lines and the defects, the coil was placed perpendicular to the defects. Fig. 7 depicts the sketch of the sample. To give the reader a better idea of the sample's geometry, the surface and subsurface defect depths are shown in Table I.

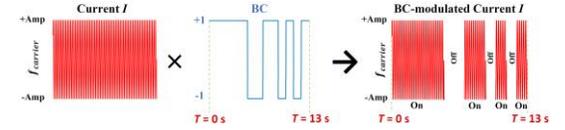

Fig. 4. Employed BC modulation on eddy current excitation.

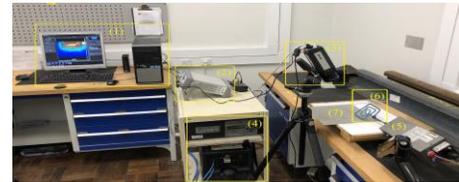

Fig.5. Experimental setup: (1) computer with software, (2) signal generator, (3) IR camera, (4) EasyHeat induction heating system, (5) work head, (6) coil, (7) sample under test.

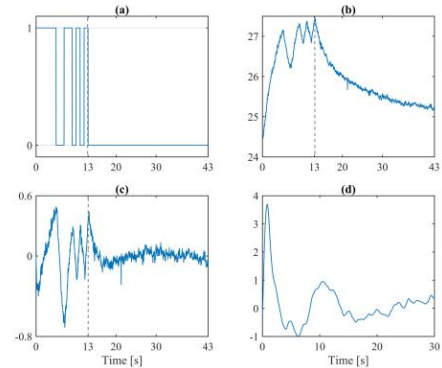

Fig. 6. (a) applied Barker Code signal; a single pixel reshape (b) of the raw acquired signal; (c) same signal as in (b), but after having performed the step-heating contribution removal; (d) signal after Pulse-compression.

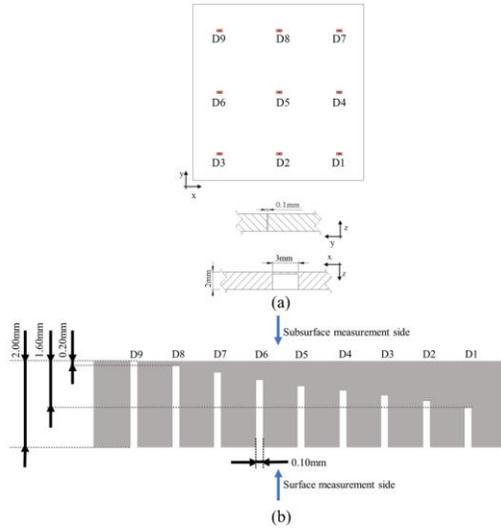

Fig. 7. Sketch of the sample: (a) plane view;(b) cross section sketch under different inspection modes.

TABLE I
SUBSURFACE AND SURFACE DEFECT DEPTHS (MM)

| Defect No. Location | D1 | D2 | D3 | D5 | D5 | D6 | D7 | D8 |
|---|---|---|---|---|---|---|---|---|
| Surface | 0.40 | 0.60 | 0.80 | 1.00 | 1.20 | 1.40 | 1.60 | 1.80 |
| Subsurface | 1.60 | 1.40 | 1.20 | 1.00 | 0.80 | 0.60 | 0.40 | 0.20 |

## IV. RESULTS AND ANALYSIS

### A. Subsurface defect location and detection techniques

After the mathematic discussion in Section II-B, two pattern recognition techniques, K-PCA and LRS, were applied and compared in this work for defect detection. The reconstruction images of subsurface defects of D8 with 0.20 mm depth, D7 with 0.40 mm depth and D4 with 1.00 mm depth are shown in Fig. 8 to Fig. 10 respectively. Before further analysis of the reconstructed images, it should be noted that the EC skin depth at the frequency of 270 kHz in Al is approximately 0.22 mm, therefore it is expected that only for D8 (with depth equal to 0.20 mm) the main contribution will be given by Joule's heating, while the thermal diffusion will be the dominant phenomenon for the deeper subsurface defects. In Fig. 8, a defective area is observed at the bottom of the crack in K-PCA1 to K-PCA3 and LRS2 to LRS4 images due to the increased EC density caused by diversion of induced current lines around the edges of the crack, since the eddy currents will always follow the path of minimum resistance. Hence, in non-defective area, EC lines are distributed evenly on the surface adjacent to the coil in a depth equal to the standard penetration depth. When a discontinuity exists, it interrupts or deviates the EC lines. Aggregation points will be made at the two crack ends, generating hot spots in thermal images.

The quantitative comparison between K-PCA and LRS is conducted in terms of SNR according to Eq. (15). The maximum SNR values obtained for defects D8 to D4 by K-PCA and LRS are shown in Table II and Table III. It is noted that LRS method provides better robustness than K-PCA in the first four components, suggesting that LRS can be applied for locating the defect when defect's depth is lower than 0.20 mm. However, K-PCA provides higher SNR maxima than LRS for D8 and D7. It can be also noted that the maximum SNR values of defects obtained by K-PCA and LRS show monotonic relationship with defect depths as illustrated in Fig. 11. Due to direct interaction of EC, defect D8 is identified with significantly higher SNR than other defects.

The improvement gained by using K-PCA and LRS are visible if the best thermal images obtained just after performing the PuC algorithm are shown. For this reason, the thermal images having the maximum SNR values for the investigated defects after PuC are depicted in Fig 12. It can be noted that the defect signature is less significant if compared with the K-PCA and LRS reconstruction images in Fig.8, Fig.9, Fig.10. For instance, the hot spot showed in Fig. 8, which is the signature of subsurface defect with 0.2mm depth, is clearer compared with best thermal image of D8 shown in Fig. 12. In addition, the best thermal images of D7 and D4 present lower signature of defects, while K-PCA3 images in Fig. 9 and Fig. 10 can reveal the location of defect.

To summarize, although the relationship of SNR and depth remains monotonic, the sensitivity is reduced when the defect is beyond the skin depth. Starting from these results, the defective pixels and non-defective pixels' time-trend are analyzed for further study in the next Section.

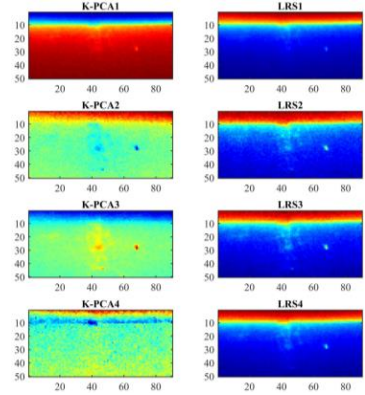

Fig. 8. Detected area using K-PCA and LRS of D8 with 0.20 mm depth in subsurface mode.

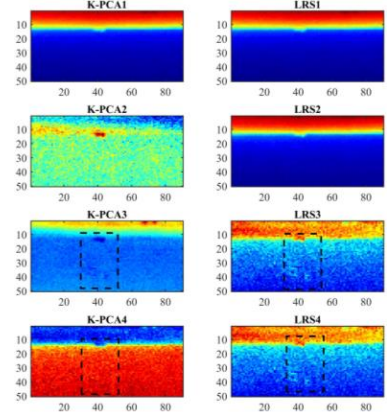

Fig. 9. Detected area using K-PCA and LRS of D7 with 0.40 mm depth in subsurface mode.

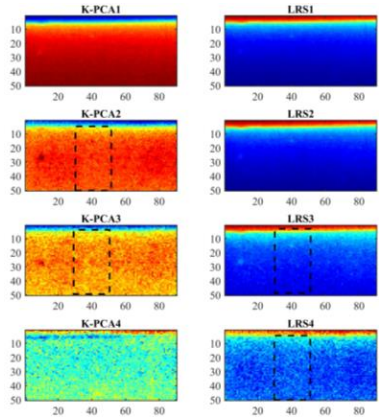

Fig. 10. Detected area using K-PCA and LRS of D4 with 1.00 mm depth in subsurface mode.

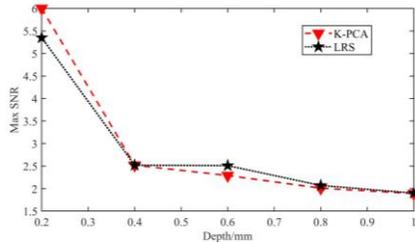

Fig. 11. Relationship between maximum SNR and defect depth in K-PCA and LRS.

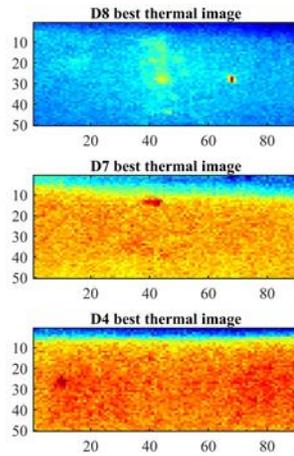

Fig. 12. Best thermal images obtained after Pulse-compression.

TABLE II
SNR VALUES FOR K-PCA

| Defect No. | PC1 | PC2 | PC3 | PC4 |
|---|---|---|---|---|
| D8 (0.20mm) | 3.99 | 5.34 | 6.00 | 0.05 |
| D7 (0.40mm) | 2.34 | 0.82 | 2.54 | 1.06 |
| D6 (0.60mm) | 2.29 | 2.16 | 1.79 | 1.37 |
| D5 (0.80mm) | 2.01 | 0.54 | 1.01 | 0.69 |
| D4 (1.00mm) | 1.89 | 0.48 | 1.02 | 0.94 |

TABLE III
SNR VALUES FOR LRS

| Defect No. | LRS1 | LRS2 | LRS3 | LRS4 |
|---|---|---|---|---|
| D8 (0.20mm) | 4.27 | 4.82 | 5.35 | 5.00 |
| D7 (0.40mm) | 2.40 | 2.29 | 2.52 | 2.25 |
| D6 (0.60mm) | 2.42 | 2.51 | 2.34 | 2.35 |
| D5 (0.80mm) | 2.03 | 1.88 | 2.07 | 1.41 |
| D4 (1.00mm) | 1.58 | 1.90 | 1.34 | 0.12 |

## B. Comparison study of depth evaluation for surface and subsurface defects using crossing point feature

As discussed in the previous section, the pattern of subsurface and surface defects was enhanced through K-PCA and LRS techniques. In this section, the impulse responses of defective and non-defective areas were selected for crossing point feature calculation. The defective pixels were manually selected as a 10 × 1 pixels area, while a non-defective area having the same size was selected 6 pixels away from the right side of defective area. Then, the impulse responses were averaged over the selected pixels and the crossing point of the two curves was obtained accordingly. The proposed crossing point feature was previously validated for evaluation of delamination depth in a Carbon fiber reinforced polymer (CFRP) material in reflection and transmission mode using ECPuCT [9].

It can be observed from Fig. 13 and Fig. 14 that the amplitude of the impulse response of the defective area is larger than that of non-defective one. The relative peak difference between defective and non-defective in subsurface mode is smaller than that of surface mode due to lower SNR values for subsurface defects.

For subsurface defects' depth evaluation, it is observed from Fig. 13 that all defect and non-defect impulse responses have two crossing points, as highlighted in the same figure. The two crossing points can be interpreted by the defect's interaction with EC and the thermal wave. Thus, the first crossing point suggests the dominant Joule heating effect, while the second one can be associated to the dominant thermal diffusion. In case of D8 in Fig. 13, the presence of two crossing points in both heating and cooling stages can be related to the fact that defect's depth almost coincides with the skin depth. In fact, the theoretical skin depth for Aluminum is 0.22 mm, so both phenomena are simultaneously significant as mentioned in Part A of the current Section. Thus, the following general conclusion can be inferred: if the first crossing point is not at the starting point as it shown for D8 and D9 in Fig. 13, the defect is within the skin depth and it is detectable due to the EC interaction difference between defect and non-defect areas. On the other hand, if two crossing exist, one just as a superposition happening at the beginning and one after the peak, *i.e.* in cooling stage, this indicates that the defect is beyond the skin depth, see for example D4-D7 in Fig 13. In other words, the EC difference between defect and non-defect areas does not exist or is considerably low and strictly requires the thermal waves to interact to be detected.

Fig. 14 shows that two crossing points are found due to Joule heating and thermal diffusion for surface defects, thus one during the heating stage and the other happening during the cooling stage. Furthermore, it is also found that all the surface defects' first crossing points are not exactly at the starting point, and this is due to the EC difference between defect and non-defect area, see Fig 14. It should be pointed out that the first crossing point of surface defect is not reliable for a faithfull depth evaluation. This is because there is a fast interaction of EC with surface defects, provoking the first crossing points of D1 to D8 in Fig. 14 happening all during the first 15 frames, therefore difficult to be resolved in the time domain. In addition, all the impulse responses of surface defects reach the peak value at the same time, which is shown in Fig. 16. Furthermore, Fig. 16 highlights that two stages can be identified in impulse response for surface defects: heating and cooling responses. The heating response of all the surface defects is due to interaction with EC, which generates different response delay such that deeper defects are heated slower than shallower ones. Instead, mentioned delay is more significant during the cooling stage, showing that deeper defects cool down slower than shallower ones. Thus, the crossing point in the cooling stage can quantify the defect depth in surface mode as showed in Fig. 14 and Fig. 15(b). In addition, in subsurface mode, the crossing point time instant is found within a certain period, *i.e.* from frame 154 to frame number 174.

Based on the above discussion for selecting approriate crossing points for depth evaluation, the finds can be generalised to the following statements:
1) The first crossing point can be used for subsurface defects, provided that the defect is within the skin depth; furthermore the determing criteria for defect within the skin depth is based on the first actual crossing point, thus excluding the crossing at the starting point due to EC difference;
2) The second crossing point should be used if the defect is located beyond the skin depth.
3) For surface defects, the second crossing point for depth evaluation shall always be used due to fast EC interaction.

As showed in Fig. 15(a) and Fig.15(b), the crossing point feature in subsurface surface modes shows a monotonic relationship with defects' depth, provided that an appropriate crossing point is considered as for the above discussion. The crossing points in defects D4 to D7 (deeper ones) showed in Fig 13 are in the cooling stage of the impulse responses, which corresponds to the depth of the defect and the heat transfer phenomena, *i.e.* deeper defects need longer time to be detected. Note also that the detection capability of ECPuCT for subsurface defects is limited to 1.00 mm depth, a fact that is due to a significant thermal conductivity and fast diffusion of heat in Al material.

The new proposed features can also be compared with previous work reported in [19]. It is known that the defect depth can be assessed by calculating the time instant at which the SNR is maximum. Fig. 17 depicts the time at which the maximum SNR is found for a given defect against the defect depth. The mentioned relation shows a monotonic trend with defect depth, but it has less sensitivity toward increasing depths compared with the crossing point feature. The same shortcoming can be observed in Fig. 11 as well, *i.e.* SNR feature has less sensitivity for deeper defects. Based on the current research on Al and previously on a CFRP sample [9], the proposed crossing point of impulse response feature from defective and non-defective areas is then validated.

Through above discussion and analysis, it has been proved that the time instant (or frame number) of crossing points between impulse responses from defective and non-defective areas can help to estimate the defect depth. It is also possible to determine whether the defect is within the skin depth of EC or not, based on the crossing point position.

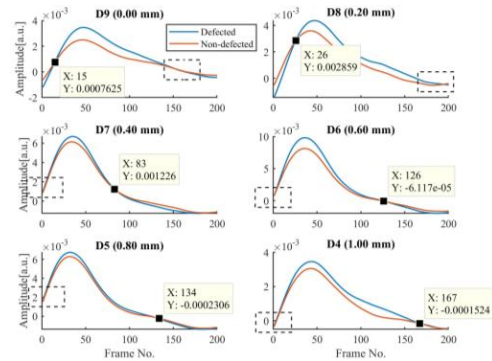
Fig. 13. Crossing point feature of subsurface defect depth evaluation.

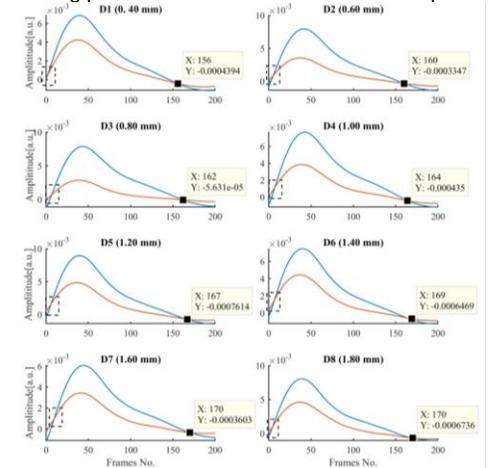
Fig. 14. Crossing point feature of surface defect depth evaluation.

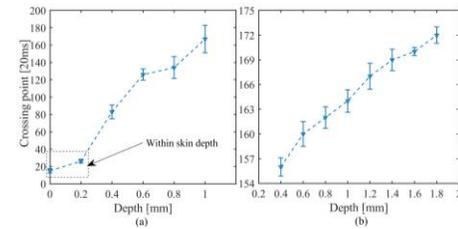
Fig. 15. Error bar plot of crossing point feature in: (a) subsurface defect, (b) surface defect.

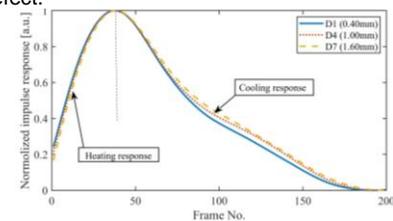
Fig. 16. Normalized impulse responses of surface defect of D1, D4 and D7.

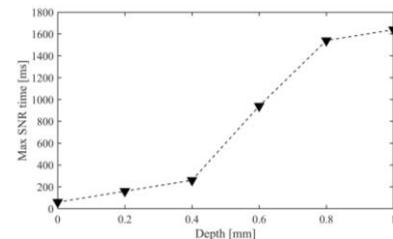
Fig. 17. Feature of maximum SNR time for subsurface defect depth evaluation.

## C. Comparison study between composites and Aluminum material with proposed feature.

As mentioned before, ECPuCT system was also applied on CFRP to quantify the depth of artificial delamination buried from 0.46 mm to 2.30 mm from the inspected surface. This Section compares the difference between the impulse responses of CFRP and Al to better illustrate the physics behind.

It can be observed from Fig. 18 that the impulse responses of CFRP and Al have different behaviors. With respect to the investigated CFRP sample, Al heats up and cools down faster due to its higher thermal conductivity. During the cooling stage, the impulse response of CFRP shows significant reduction of cooling rate around the $100^{th}$ frame, i.e. $t = 2\ s$. This can be interpreted as the diffusion of the heat from the inner volume to the surface, leading to a reduction in the cooling rate in later response. The skin depth of CFRP is higher than that of Al material due to lower thermal conductivity. In particular, for the same $f_{carrier}$ value mentioned in Section III, the skin depth in CFRP is higher than the thickness of the sample itself, thus the heating mode can be considered as volumetric. On the other hand, the heat transfer in the Al sample happens mainly close to the surface because the skin depth is significantly lower in value than the sample thickness. We believe that the volumetric heating mode causes the difference in the response curves, *i.e.* slower cooling rate in CFRP with respect to Al sample, as illustrated in Fig. 18(a) and Fig. 18(b). This hypothesis is also supported by other experimental findings. For instance, as highlighted in Fig 17, the peak time of Al sample was found slightly earlier than that of CFRP samples. Further, surface defects with different heights have the same peak time shown in Fig. 16, a fact that is in line with previous work making use of PEC [20]. In addition, the significant difference in the cooling trend of the impulse responses of Al and CFPR samples is shown in Fig 18. Further quantitative analysis can also be conducted in line with thermographic signal reconstruction (TSR) and other transient thermal responses [21].

Fig. 19 shows the depth evaluation of subsurface defects in Al and CFRP by crossing point feature. Due to the volumetric heating, this feature in CFRP has stronger linearity when plotted versus defect's depth with respect to the same analysis carried out for the Al sample. According to Section IV B, the crossing point on the left side of the peak or near the peak indicates the dominance of Joule heating. For the CFRP sample, the estimated skin depth is 1.84mm, so most of the defects lie in the skin depth range, as shown in Fig 19(a), though it must be also considered that EC are induced even deeper, and indeed the crossing point feature is quite linear for all the defects. It is also observed that the crossing point feature in Al and CFRP for defects within the skin depths exhibits different linear slopes compared with the feature in thermal diffusion stage for subsurface defects.

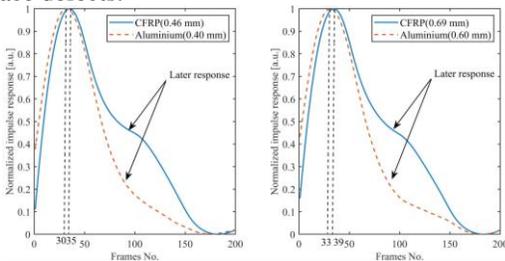

Fig. 18. Comparison of normalized impulse response of subsurface defects: (a) CFRP sample with defect depth of 0.46mm, Aluminium sample with defect depth of 0.40mm;(b) CFRP sample with defect depth of 0.69mm, Aluminium sample with defect depth of 0.60mm.

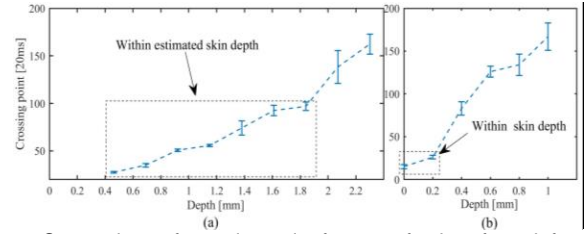

Fig. 19. Comparison of crossing point feature of subsurface defects: (a) CFRP sample with defect depths of 0.46 mm to 2.30 mm ;(b) Aluminium sample with defect depths of 0.00 mm to 1.00 mm.

## V. SUMMARY AND FUTURE WORK

This work presented the application of ECPuCT to detect subsurface and surface defects with various depths on Al sample. The conclusions are as follows:
1) The proposed crossing point feature between defective and non-defective impulse responses has monotonic relationship with the depth of subsurface and surface defects in Al material;
2) The subsurface defects within eddy current skin depth have two crossing points. If the defect depth is smaller than or comparable to the skin depth, the first crossing point can be used for depth estimation; when defects' depth is larger than the skin depth, the first crossing point is the starting point due to no EC difference, thus only the second crossing point in the cooling stage is meaningful and it should be used for defect depth estimation;
3) Similarly, surface defects have two crossing points. Thus, the evaluation of defect depths can only be conducted in the cooling stage with proposed feature. This is because the fast interaction between EC and defect does not produce faithful impulse response within the heating stage;
4) The proposed crossing points show a good ability to evaluate defect depths, not only within the eddy current skin depths but also beyond it.

It must be stressed out that the abovementioned findings, related to the use of K-PCA and LRS applied over ECPuCT, can be extended to PEC thermography for example but also to any other stimulated thermography implying volumetric heating, provided that the hypothesis of Linearity and Time Invariance are fulfilled.

Future works will investigate the evaluation limitation of defect depths in the cooling stage in comparison with other active thermography scheme. Furthermore, next works will be focused on the analysis of different stages of impulse responses for extracting multiple properties of the defects, such as their width and thickness. It would be also interesting to compare the defect depth evaluation capability of ECPuCT with that achievable by using EC testing on a given benchmark sample, together with machine learning and feature extraction approaches [21].